\documentclass{jpp}

\usepackage{graphicx}
\usepackage{amsmath}
\usepackage{amsfonts}
\usepackage{amssymb}
\usepackage{graphicx}
\usepackage{mathrsfs}

\usepackage{epsfig}
\usepackage{natbib}

\ifCUPmtlplainloaded \else
  \checkfont{eurm10}
  \iffontfound
    \IfFileExists{upmath.sty}
      {\typeout{^^JFound AMS Euler Roman fonts on the system,
                   using the 'upmath' package.^^J}%
       \usepackage{upmath}}
      {\typeout{^^JFound AMS Euler Roman fonts on the system, but you
                   dont seem to have the}%
       \typeout{'upmath' package installed. JPP.cls can take advantage
                 of these fonts, if you use 'upmath' package.^^J}%
      }
  \else
  \fi
\fi

\ifCUPmtlplainloaded \else
  \checkfont{msam10}
  \iffontfound
    \IfFileExists{amssymb.sty}
      {\typeout{^^JFound AMS Symbol fonts on the system, using the
                'amssymb' package.^^J}%
       \usepackage{amssymb}%

      }{}
  \fi
\fi

\ifCUPmtlplainloaded \else
  \IfFileExists{amsbsy.sty}
    {\typeout{^^JFound the 'amsbsy' package on the system, using it.^^J}%
     \usepackage{amsbsy}}
    {\providecommand\boldsymbol[1]{\mbox{\boldmath $##1$}}}
\fi

\newsavebox{\astrutbox}
\sbox{\astrutbox}{\rule[-5pt]{0pt}{20pt}}

\title[Towards Quantum Gravity measurement by cold atoms]{Towards Quantum Gravity Measurement by Cold Atoms}

\author[M. M. Dos Santos, T. Oniga, A. S. Mcleman, M. Caldwell, C.H.-T. Wang]%
{M\ls A\ls R\ls C\ls I\ls L\ls I\ls O\ls \ns M. D\ls O\ls S\ns  S\ls A\ls N\ls T\ls O\ls S$^{1,2}$%
  \thanks{Email address for correspondence: c.wang@abdn.ac.uk},\ns
T\ls E\ls O\ls D\ls O\ls R\ls A\ns  O\ls N\ls I\ls G\ls A$^1$,\break
A\ls N\ls D\ls R\ls E\ls W\ns S.\ns M\ls C\ls L\ls E\ls M\ls A\ls N$^1$,\ns
M\ls A\ls R\ls T\ls I\ls N\ns C\ls A\ls L\ls D\ls W\ls E\ls L\ls L$^2$  \and C\ls H\ls A\ls R\ls L\ls E\ls S\ns H.\ls -\ls T.\ns W\ls A\ls N\ls G$^{1,2}$}

\affiliation{$^1$SUPA Department of Physics, University of Aberdeen, King's College, Aberdeen AB24 3UE, UK\\[\affilskip]
$^2$STFC Rutherford Appleton Laboratory, Chilton, Didcot, Oxon OX11 0QX, UK}

\date{December 2012}
\begin{document}

\maketitle

\begin{abstract}
We propose an experiment for the measurement of gravitational effect on cold atoms by applying a one-dimensional vertically sinusoidal oscillation to the magneto-optical trap; and observe the signature of low quantum energy shift of quantum bound states as a consequence of gravitational fluctuation. To this end, we present brief details of the experiment on a BEC, and a simplistic calculation of the Gross-Pitaevskii solution using Thomas-Fermi approximation with focus on the density of the BEC, for the time-dependent perturbation.  Moreover, we calculate the power induced by quantum gravity on a generic atomic ensemble. We also address the possible challenges for the measurement of the expected results. And finally, we discuss the prospect of further developing this experiment towards measuring the effect of quantum spacetime fluctuations on cold atoms. 
\end{abstract}


%05.40.-a Fluctuation phenomena, random processes, noise, and Brownian motion 
%04.60.Bc Phenomenology of quantum gravity 
%67.85.-d Ultracold gases, trapped gases 
%03.75.-b Matter waves 

\section{Introduction}
 The use of cold atoms allows new ways of testing fundamental quantum phenomena \citep{Anderson}. Among them is the elusive quantum gravity and its effect, which for years has avoided been understood and detected. Despite all the unsuccessful attempts, there have been suggestion that the low energy signature of quantum gravity may be detectable \citep{Charles12}. According to the authors; and inspired by the success from the Lamb shift measurement in QED \citep[see][]{Lamb55, Bethe50}, they believe there is a possibility of observing the energy shifting of the quantum bound state coupled to a quantum field under vacuum fluctuation; similar to that discovered by Lamb. Incidentally, due to the fact that the lowest energy of quantum gravity is the zero-point energy of vacuum, then by manipulating cold atoms or a BEC center of mass, formed in the lowest state, the energy shift may be amplified to the point of observation.\\

In this paper we will present an experiment to investigate the behaviour of the BEC under such conditions, as required by the gravitational Lamb shift. Initially, a brief introduction about the gravitational Lamb shift mentioned above will be presented; then, we will describe the experiment to be cared out in the near future; and  after, a brief analysis of the BEC behaviour and calculations of the Gross-Pitaevskii  equation solution under the harmonic perturbation focused on the density of the BEC. Moreover, we will calculate the general quantum mechanical case of an ensemble of atoms under the same harmonic perturbation. From there, we were able to observed that the ensemble atoms does indeed absorbs energy from the gravitational induced perturbation. Finally, we present the technical difficulties that may compromise the experiment in general, and also underline the results as well as the uncertainties of the experiment in the discussion.

\section{Gravitational Lamb Shift}

The idea of a gravitational Lamb shift according to \citep{Charles12} is based on the same principle used to determine the QED Lamb shift used for the Hydrogen atom. The paper used similar semiclassical derivation of the Lamb shift as derived by Welton theory \citep{Welton97}. From such derivation they found that the energy shift of the condensate in a harmonic potential would be zero, unless the center of mass of the BEC is displaced from its original position. In that case, they found that the expected potential energy $V$ is given by 

\begin{equation}
\langle \Delta V \rangle \approx \frac{16}{27\pi} \frac{m^{3}}{m_{P}^{2}} \omega^{2}L^{2} 
\end{equation}
with $m, m_{P},L$ being the atomic mass, Planck mass, and trap vertical length. Moreover, if they were using a frequency of $\omega = 2 kHz$ with the number of atoms in the condensate equal $N=10^{6}$, then the energy ratio $\Delta E /\hslash \omega $ would have increased to 0.5 $\%$. Detailed information on their derivation of the gravitational Lamb shift is found in \citep[see][]{ Charles12}.

\section{Proposed experiment setup}
In this paper we propose an experiment for measuring quantum gravity based on cold atoms. To obtain a good guidance of the phenomenon, we will consider a simplified one-dimensional harmonic oscillation for the cold atoms and BEC. The trap used is a normal anisotropic spherical shape MOT. There is an illustration of the main equipments used in the experiment on fig.(1).
The aim of the experiment is to achieve a reasonable displacement of the atom cloud or BEC center of mass from its origin, with a specific frequency. Such displacement, in our set up, provides a mechanical vertical sinusoidal oscillation of the MOT assemble due to an electro-acoustic vibration generator. \\

\begin{figure}
\centering
\includegraphics[scale=0.35]{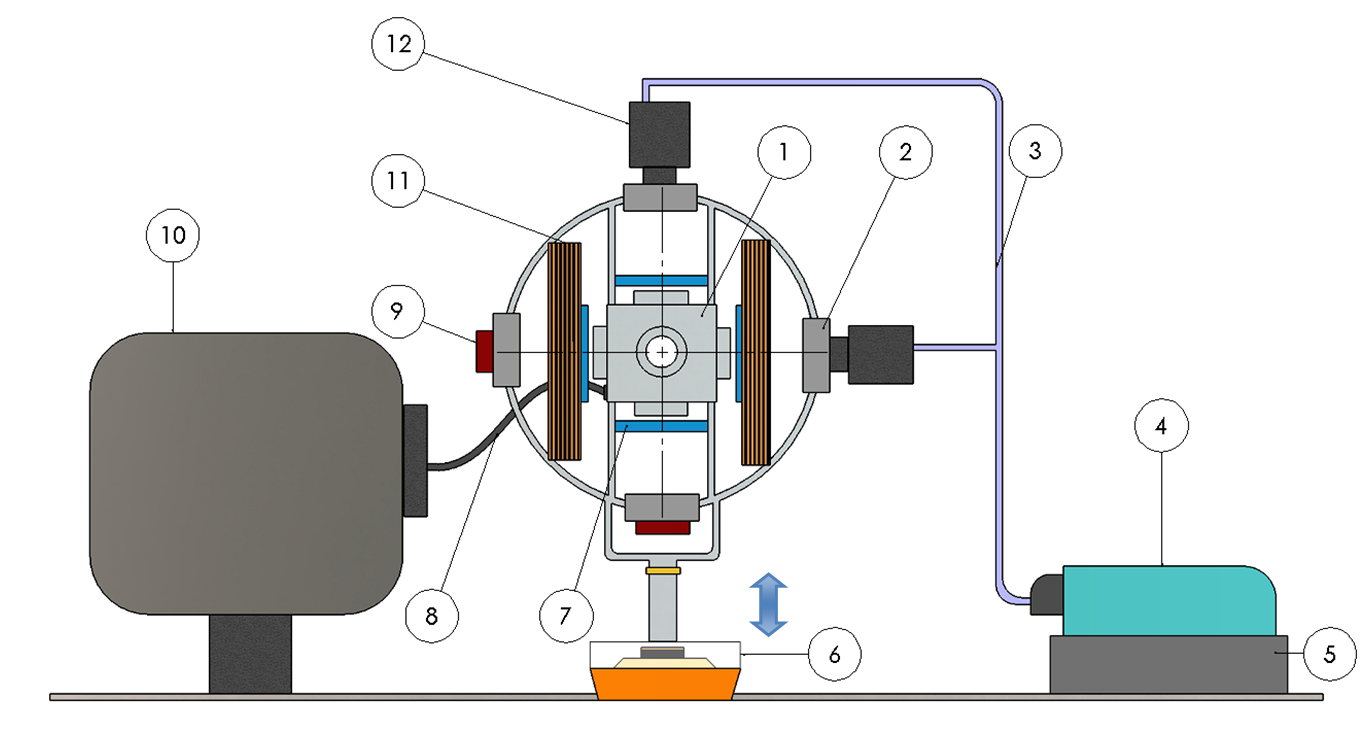}
\caption{Experiment setup, 1- trap, 2- quarter wave plate, 3- fibre optics, 4- tuneable laser, 5- electro-acoustic vibrator drive, 6-electro-acoustic vabration generator, 7- TOP maginetic coils, 8-  flexible vacuum pipe, 9- mirror, 10- ion pump, 11- trap magnetic coils (MOT), 12- telescope}\label{fig1}
\end{figure}

The trap displacement is up to 1 cm (susceptible to change), giving the sine wave an amplitude of the same value. Incidentally, we assume that all the necessary conditions for the creation of BEC are met as well as the structure of the anisotropic harmonic trap is maintained under constant oscillations, that is, constant at the origin and during the oscillation of the system. The sinusoidal varying system will allow the formation of a harmonic perturbation of the metric due to the oscillating potential, which corresponds to a function of space and time in the direction of coordinate $z$ given by
\begin{equation}
z(t)=a \sin(\Omega t)
\end{equation}
where $a$ is the amplitude and $\Omega=2\pi f$ the angular frequency of the oscillation.

The acceleration is given by
\begin{equation}
A(t) = -\Omega^{2}a \sin(\Omega t) .
\end{equation}

This gives rise to the time dependent potential perturbation as follows: 
\begin{equation}
V(z,t) = m A(t)z
\end{equation}
using the BEC or atomic ensemble mass $m$. For this particular experiment we assume the phase to be zero and the frequency on the order of 1-2 kHz. The atom vapour species that we will be using is the isotope of Rubidium Rb87. The initial number of atoms  in vapour will be of order $10^{12}$ with the condensate having about $2000$ to $10^6$ atoms. In this  case the interaction will be repulsive, that is, s-wave scattering length $a_{s}$ is positive. Within our 2.5 $cm{^{3}}$ trap, we would expect $2\times 10^{3}$ -- $2\times 10^{4}$ atoms after the spatial distribution forming the isotropic thermal equilibrium condensate in the ground state (BEC). Then, by increasing the number of atoms in the trap to form a larger BEC,  we will be able to observe if the corresponding effect of gravitational lamb shift will be proportional to that predicted by the theory. \\

 The 3D harmonic potential (i.e. TOP time orbiting potential of 7.5 kHz) generated by the magnetic coils are meant to be relatively symmetrically constant during the system oscillation. The time-dependent perturbation created  by the vibration generator would allow the necessary conditions for the gravitational effect predicted to be exposed and detected. This vibration generator provides $5N$ of vertical force, and an oscillating frequency that varies from 1Hz to 15kHz, once controlled by the drive.\\
 
  Now, from the schematic above, it is worth mention that we did not show the RF bias magnetic field, external back side window of CCD camera, the front TOP magnetics coils , control systems, detection and the remaining Ioffee-Pritchard configuration mechanisms. Incidentally, regarding the detection mechanism, we will be using a probing laser and a CCD camera to measure the level of energy, density and the decoherence of the condensate or atomic cloud.

\section{Harmonic Perturbation on BEC}

We used the Gross-Pitaevskii equation \eqref{4.1} to study the behaviour of the BEC under the Harmonic perturbation,

\begin{align}
i\hslash\dfrac{\partial \psi(X,t)}{\partial t} = -\dfrac{\hslash^{2}}{2m}\bigtriangledown^{2}\psi(X,t)+V(X,t)_{T}\psi(X,t)+
\frac{4\pi\hslash^{2}a_{s}}{m}|\psi(X,t)|^{2}\psi(X,t) .\label{4.1}
\end{align}

There we have used $V(X,t)_{T}$ as the sum of the potential $V(z,t)$ to the Anisotropic potential $V(X,t)_{trap}$, and $X =x,y,z$ coordinates:\\

\begin{align}
 V(X,t)_{T}=\dfrac{1}{2}m_{a}\omega^{2}_{x}x^{2}+\dfrac{1}{2}m_{a}\omega^{2}_{y}y^{2}+\dfrac{1}{2}m_{a}\omega^{2}_{z}z^{2}+mA(t)z .
\end{align}

For a simplified analyse, we consider the numerical solution of the equation \eqref{4.1} using Thomas-Fermi Approximation. For this purpose, it is necessary that $V(X,t)_{T}$ is less than the chemical potential $\mu$, in this paper we neglect the trap frequencies (i.e. the radial and axial frequencies $\omega_{x}=\omega_{y}$, and $\omega_{z}$ ). Here, $m_{a}$ is the atomic mass of Rb87, $m$ the total mass of the BEC which will be $N_{0}m_{a}$, with $N_{0}$ being the number of atoms in zero state, N the number of atoms. We also use an amplitude for the mechanical oscillation of $1cm$ for this particular case; the s-wave scattering length $ a_{s} = 5.28nm$, as for the Rb 87. Incidentally, an observation of the effect of our potential using GPE solution towards the density of the BEC was obtained. Here, it was noticed that our potential generates a variational solution of GPE, and was acquired using Thomas-Fermi Approximation. Fig.(2), illustrated the behaviour of the matter wave density, as a function of time. There, it was noticed that as the mechanical oscillation evolved, the ratio of the density of Thomas-Fermi density $n_{TF}$  with $n$ increased quite rapidly, and then collapsed to an imaginary solution of the wave function continuously. The Thomas-Fermi density $n_{TF}$  is given by\citep{pethick02}
\begin{equation}
n(z)=n_{TF}=(\mu-V(z,t))/g
\end{equation}\\
where $g=4\pi\hslash^{2}a_{s}/m$, $\mu>V(X,t)$, with $a_{s}= 5nm$, and mass of Rb87 BEC,  $\mu \approx ng$ with $n =\mu(0)/k_{B}T_{c}\approx 0.3$ \citep{Dalfovo99} . The mechanical perturbation creates a rising effect on the total density of the condensate. We noticed that the effect forces the BEC to go beyond the Thomas-Fermi regime, giving negative density and therefore, forcing the solution of the wave function to be complex. From the graphic on the left of fig.(2), the density ratio increases to $0.5 \% $ above 1 with $N_{0}=1000$, amplitude 1 cm and frequency 1kHz; Whereas, for the graphic on the right, with the difference only in $N_{0}=10^{6}$, we observed and increase of $50 \%$. These particular increases in the density ratio indicates that the Harmonic mechanical perturbation does affect the energy level to certain extent. However, a thorough analyse with different methods such as the Bogoliubov theory may be consider in future papers.  \\

\begin{figure}
\centering
\includegraphics[scale=0.4]{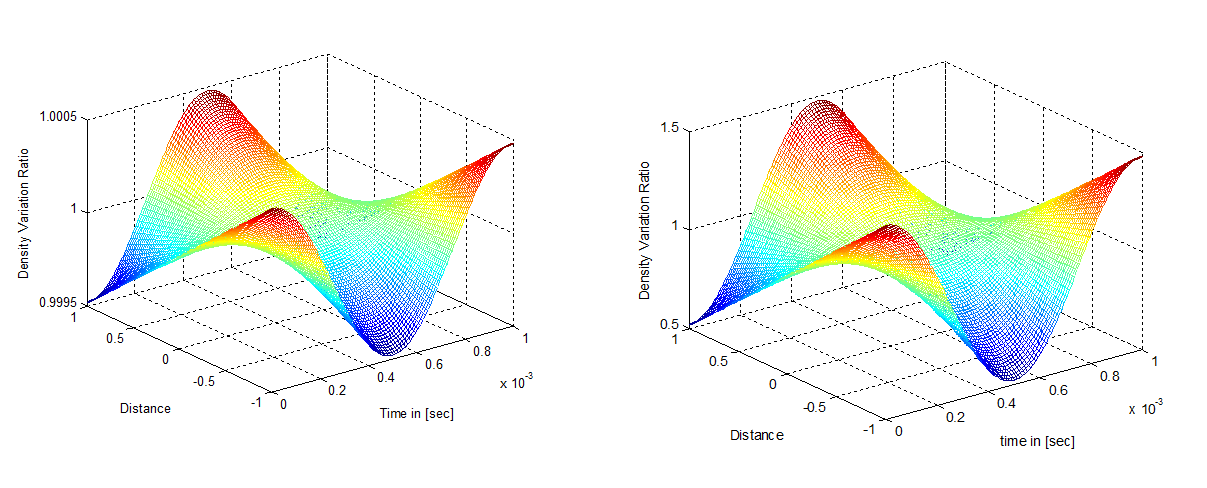}
\caption{Density ratio variation with 1kHz, 1 cm amplitude, $N_{0}=1000$ (left), and $N_{0}=10^{6}$ (Right)}\label{fig}
\end{figure}

\section{Gravitational Power Absorption}

Here, we start by considering a one-dimensional harmonic oscillator in an initial state $|i\rangle$. The harmonic oscillator is then perturbed by a small time-dependent potential $V(x,t) = mA(t)x$ with a general (broadband) time dependent acceleration $A(t)$. The probability of finding the oscillator in the state $|n\rangle$ at a later time can be found using perturbation theory. 

We therefore consider the time-dependent Hamiltonian given as,

\begin{equation}\label{darp}\large{\hat{H} = \hat{H}_0 + V(x,t)}\end{equation}

where

\begin{equation}\label{vax}\large{V(x,t) = mA(t)x} .\end{equation}

The generic time-dependent Hamiltonian in equation \eqref{darp} does not have a typical analytical solution. Using perturbative analysis, the basis coefficients $c_n$(t) can be expanded in powers of the interaction.

\begin{equation}\large{C_n (t) = c_n^{(0)} + c_n^{(1)} (t) + c_n^{(2)} (t) + \cdots .}\end{equation}

For the generic potential, different cases of $A(t)$ can be considered. For example, a harmonic potential can be given by

\begin{equation}\large{A(t) = A_0 \sin(\Omega t)} .\end{equation}

It follows that

\begin{equation}\large{V_{ni} (t^{\prime}) = m\langle n | V(x,t) | i \rangle} .\end{equation}

Substituting equation \eqref{vax} into the above gives

{\large
\begin{eqnarray}
\label{vatx}
V_{ni} (t^{\prime}) 
&=&
m A(t) \langle n |  x | i \rangle .
\end{eqnarray}
}

Since

{\large
\begin{eqnarray}
\langle n | x | i \rangle 
&=& 
\sqrt{\frac{\hbar}{2mw}} \langle n | \hat{a} + \hat{a}^{\dagger} | i \rangle
\nonumber
\\
&=&
\label{delta}\sqrt{\frac{\hbar}{2mw}}  (\sqrt{i + 1}\delta_{n, i + 1} + \sqrt{i}\delta_{n, i - 1}) 
\end{eqnarray}
}
we can calculate the first order perturbation  coefficient as follows: 

\begin{equation}\label{58}\large{ c^{(1)}_{ni}(t) = - \frac{i}{\hbar} \int_{t_0}^{t} dt^{\prime} e^{iw_{ni}t^{\prime}} V_{ni} (t^{\prime})} \end{equation}

Substituting equation \eqref{vatx} into this equation yields

\begin{equation}\label{inttt}\large{ c^{(1)}_{ni}(t) = - \frac{i m}{\hbar} \langle n | x | i \rangle \tilde{A}_{t}(\omega_{ni}) } \end{equation}
where $\tilde{A}_{t}(\omega_{ni})$ is the Fourier transform of the complex stationary stochastic function $A(t)$, given by

\begin{equation}\label{ghost}\large{ \tilde{A}_{t}(\omega_{ni}) = \int_{0}^{t} d t^{\prime} e^{i \omega_{ni} t^{\prime}} A(t^{\prime})} . \end{equation}

The mean power spectral density of this integral is normally given by

\begin{equation}\label{tilde}\large{ S_t(\omega_{ni}) = \frac{1}{t} \langle \tilde{A}_{t}^{*}(\omega_{ni}) \tilde{A}_{t}(\omega_{ni}) \rangle }\end{equation}
where $\langle \tilde{A}_{t}^{*}(\omega_{ni}) \tilde{A}_{t}(\omega_{ni}) \rangle$ denotes the statistical average. Additionally, It is known that

\begin{equation}\large{ S(\omega_{ni}) = \lim_{t \rightarrow \infty} S_{t}(\omega_{ni})}\end{equation}

And the transition probability can be found using \eqref{inttt} to be

\begin{equation}\large{ P_{ni} = \lvert c_{ni}^{(1)}|^{2} = \rvert \frac{-i m}{\hbar} \langle n | x | i \rangle |^{2} \langle \tilde{A}_{t}^{*}(\omega_{ni}) \tilde{A}_{t}(\omega_{ni}) \rangle . }\end{equation}

Substituting the result of equation \eqref{tilde} into the above, we have

\begin{equation}\label{prob}\large{ P_{ni} = \lvert c_{ni}^{(1)}|^{2} = \frac{m^2}{\hbar^{2}} \rvert  \langle n | x | i \rangle |^{2} S(\omega_{ni}) t}\end{equation}

From this, dividing the transition probability by time gives the transition rate

\begin{equation}\label{Probs}\large{ T_{i \rightarrow n} = \frac{P_{i \rightarrow n}}{t} = \lvert c_{ni}^{(1)}|^{2} = \frac{m^2}{\hbar^{2}} \rvert  \langle n | x | i \rangle |^{2} S(\omega_{ni}) } .\end{equation}

Therefore from equations \eqref{delta} and \eqref{Probs} the transition rate becomes

{\large
\begin{eqnarray}
T_{i \rightarrow n} 
&=& 
\frac{m}{\hbar^{2}} S(\omega_{ni}) \frac{\hbar}{2 \omega} \left[ (i + 1)\delta_{n, i + 1} + i\delta_{n, i-1} \right]
\nonumber
\\
&=&
\frac{m S(\omega_{ni})}{2\hbar \omega} \left[ (i + 1)\delta_{n, i + 1} + i\delta_{n, i-1} \right] .
\end{eqnarray}
}

This gives rise to the total absorption power for the initial state $| i \rangle$ given by

{\large
\begin{eqnarray}
{\cal P}_{i \rightarrow all} 
%&=& 
%\sum_{n} (E_{n} - E_{i}) T_{i \rightarrow n}
%\nonumber
%\\
&=&
\frac{m\hbar \omega}{2\hbar \omega} \left[ (i + 1)S(\omega) \right] - \frac{m\hbar \omega}{2 \hbar \omega} \left[iS(-\omega) \right]
\nonumber
\end{eqnarray}
}
that is,
\begin{equation}\label{sw}\large{{\cal P} = \frac{m S(\omega)}{2}}\end{equation}
since  $S(\omega) = S(-\omega)$ is an even function. Here we have dropped the subsrcipt of $P$ since this absorption power turns out to be independent of the initial state.\\

% his conclusion

From this point we conclude the theoretical investigation which was carried out into the effect of a stationary time domain signal on a one-dimensional model of an atom trap. The derived relation is both concise and general, in terms of the power spectral density of the exciting force. The power spectral density term encodes full details of a broadband excitation containing a stationary signal as well as a possible additional stochastic noise. Now for the same spectral density in (5.17) when considering a simple harmonic oscillator, can be given as

\begin{equation}
S(\omega)=\dfrac{\pi A_{0}^{2}}{2}[\delta(\omega - \Omega)+\delta (\omega +\Omega)]
\end{equation}
where $\Omega$ is the frequency induced by the electro-acoustic vibration generator, and $\omega$ the internal trap frequency (i.e. the resultant frequency due to the combination of different magnetic fields from TOP and anti-helmholtz coils, $A_0$ related to the acceleration of the shaker by $A_0 = -a\Omega^{2}$. Hence, the absorption power can be expressed as 

\begin{equation}\label{finall}\large{{\cal P} = \frac{\pi m A^{2}_{0}}{4}}\delta(\omega - \Omega) . \end{equation} \label{sw}

To obtain a definite yet simple estimate, we shall average $P$ over a uniformly spread statistical distribution of the harmonic oscillators with density $\rho(\omega)=1/\Omega$ for $|\omega-\Omega|<\Omega/2$ having a unity total probability. This yields the final expected value for the absorbed power by the atom trap as follows: 

{\large
\begin{eqnarray}
\langle {\cal P}\rangle
&=& 
\int_{\Omega-\Omega /2}^{\Omega+\Omega/2} \large{{\cal P}}(\omega) \rho(\omega)d\omega
\nonumber
\\[5pt]
&=&
\frac{\pi m a^2\Omega^3}{4} .
\end{eqnarray}
\label{sw}
}

\begin{figure}
\centering
\includegraphics[scale=0.5]{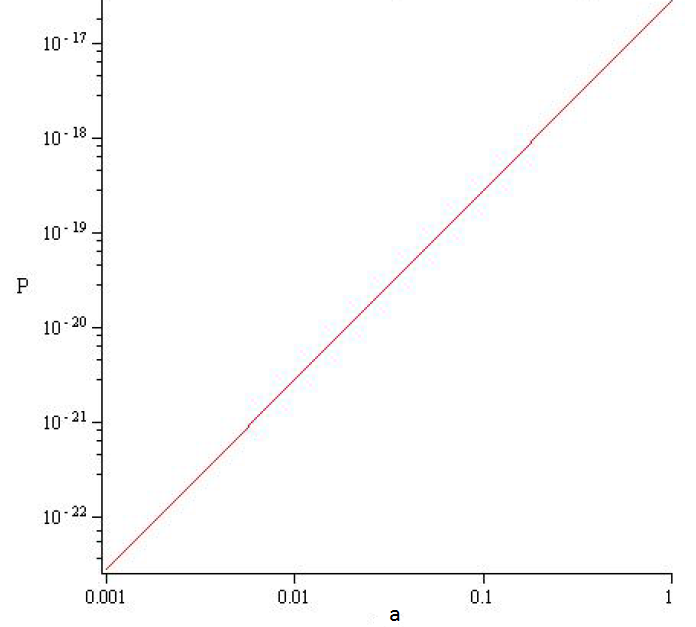}
\caption{Double-logarithimic plot of power (W) as function of $a$ (mm). }\label{fig1}
\end{figure} 

As it can be noticed, the power produced by an atom is quite large on a timescale of a second compared to $k_BT$ for a temperature of micro Kelvin and to the energy of the ground state $hf$:
\begin{eqnarray} \nonumber
&&k_bT \sim 10^{-29} \; J\\ \nonumber
&&hf \sim 10^{-31} \; J .
\end{eqnarray}
\\

This indicates that the atom trap is extremely sensitive to an extremely small perturbation; Hence, even for an extremely small perturbation the power generated would be enough to produce evaporation within a few seconds.

\section{Discussion}

From the above description, it is clear that we could indeed find the BEC and the atom cloud reacting vigorously to the harmonic perturbation induced by the electro-acoustic generator. The brief analyse in the density of the Bose-Einstein condensate in section 4, mentioned the need for a more profound understanding of this new way of exposing the condensate. We found in section 5, that by deriving from a simple harmonic oscillator, the spectral density, and transition rate of the atoms in the trap can lead to determine the total absorption power of the initial state; and finally, using the same spectral density we realized that the power induced on an atom can be extremely large as shown in fig.(3). This findings are good motivation for pursuing the experiment. \\

On the other hand, there are numerous practical issues that will need to be resolved for the experiment to be successful.  One of the problem is the influence of the trapping orbital electromagnetic field, which can potentially alter the dynamic of the condensate. For instance, we noticed that such configuration of the Quadrupole-Ioffe magnetic trap, when under the time-dependent potential of the electro-acoustic vibrator, will force the Majorana Hole to form a helix of the rotating TOP magnetic field. We need to know what combined effect it will have on the cloud or BEC.\\

For most of the situation, s-wave length $a_{s}$ is not sensitive to the external potential, however, near Feshbach resonance the condensate can switch signs (i.e. repulsive to attractive), and change the dynamics of the condensate. The collapse of BEC can happen if the critical particle number $N_{cr}=N_{0}$, that is, if during the oscillation $a_{s}$  is tuned by the trapping frequency. Another problem may arrive due to mechanical vibrations of the equipments in the movable frame in the schematic. \\

When it comes to flexibilities, one important factor is the way the increases on the energy level due to gravitational lamb shift will be measure; by using lasers off resonance, and reading the refractive index of BEC, which will differ by the presence of different energy levels, we will be able to tell to what extent the energy was increased. However, One of the constraints in this experiment is the number of atoms in the condensate necessary for the observation of gravitational Lamb shift; $10^{6}$ atoms in the condensate of Rb87 may be slightly difficult to obtain and control for a long time.\\

Finally, it is worth mention the difficulties that the experiment poses to verify the prediction of gravitational Lamb shift. However, difficult, the experiment is also viable, as long as the inertial effect will be dealt within the short lifetime of the BEC. Therefore, we feel the need to take the steps towards the development of the experiment, and study the effects to ensure that such discrepancies between gravity and quantum mechanics can be quantified and understood. And in addition to that, have experimental evidence that a clear observation of gravitational Lamb shift was achieved. 

\section{Acknowledgement}
The authors are most grateful to J. T. Mendon\c{c}a and R. Bingham for stimulating discussions.
M.M. Dos Santos is supported by Sonangol under a PhD Studentship. T. Oniga thanks the IoP Nuffield Foundation for an Undergraduate Research Bursary and A. McLeman thanks the Carnegie Trust for an Undergraduate Research Studentship. M. Caldwell and C. Wang are indebted to the EPSRC and STFC Centre for Fundamental Physics for partial support.

% susie put cite commands here, don't bother with citet etc just yet.

\bibliographystyle{jpp}

\bibliography{jpp-instructions}

\end{document}